\documentclass[aps,prl,showpacs,showkeys,twocolumn,groupedaddress]{revtex4}

\usepackage{amsmath}
\usepackage{amsfonts}
\usepackage{amssymb}
\usepackage{latexsym}
\usepackage{graphicx}
\usepackage{float}
\usepackage{wrapfig}

\newcommand{\detg}{\textrm{sdet}}
\newcommand{\tr}{\textrm{tr}}
\newcommand{\trg}{\textrm{str}}

\newcommand{\im}{\textrm{Im}}

\newcommand{\diag}{\textrm{diag}}

\newcommand{\eins}{\leavevmode\hbox{\small1\kern-3.8pt\normalsize1}}

\begin{document}
 \title{On the Eigenvalue Density of Real and Complex Wishart Correlation Matrices}
\author{Christian Recher, Mario Kieburg and Thomas Guhr}
\date{\today}
\affiliation{Fakult\"at f\"ur Physik, Universit\"at Duisburg--Essen, Lotharstra\ss e 1,
47048 Duisburg, Germany}

\begin{abstract}
  Wishart correlation matrices are the standard model for the
  statistical analysis of time series. The ensemble averaged
  eigenvalue density is of considerable practical and theoretical
  interest.  For complex time series and correlation matrices, the
  eigenvalue density is known exactly. In the real case, however, a
  fundamental mathematical obstacle made it forbidingly complicated to
  obtain exact results.  We use the supersymmetry method to fully
  circumvent this problem.  We present an exact formula for the
  eigenvalue density in the real case in terms of twofold integrals
  and finite sums.
\end{abstract}
% 05.45.Tp   Time series analysis 
% 02.50.-r     Probability theory, stochastic processes, and statistics 
% 02.50.Sk   Multivariate analysis
% 02.50.Tt    Inference methods 
% 02.20.-a    Group theory 
% 02.20.Qs   General properties, structure, and representation of Lie groups 

\maketitle

Time series analysis is an indispensable tool in the study of complex
systems with numerous applications in physics, climate research,
medicine, signal transmission, finance and many other
fields~\cite{Cha04,Mbg05,Seb03}.  A time series is an observable such
as the water level of a river, the temperature, the intensity of
transmitted radiation, the neuron activity in electroencephalography
(EEG), the price of a stock, etc., measured at usually equidistant
times $k=1,\ldots,n$. Suppose we measure $p$ such time series $M_j, \
j=1,\ldots,p$, for example, in the case of EEG, at $p$ electrodes
placed on the scalp or, in the case of temperatures, at $p$ different
locations. Our data set then consists of the $p\times n$ rectangular
matrix $M$ with entries $M_{jk}$. The time series $M_j$ are usually
real (labeled $\beta=1$), but in some applications they can be
complex ($\beta=2$). Often, one is interested in the correlations between
the time series. To estimate them, one normalizes the time series
$M_j$ to zero mean and unit variance. The correlation coefficient
between the time series $M_j$ and $M_l$ is then given as the sample
average
\begin{equation}
C_{jl} = \frac{1}{n} \sum_{k=1}^n M_{jk} M_{lk}^*
\qquad {\rm and} \qquad
C = \frac{1}{n} M M^\dagger
\label{eq1}
\end{equation}
is the correlation matrix. For real time series ($\beta=1$), the
complex conjugation is not needed and the adjoint is simply the
transpose. We notice that $C$ is a $p\times p$ real--symmetric
($\beta=1$) or Hermitean ($\beta=2$) matrix.

The eigenvalues of $C$ provide important information, see recent
examples in Refs.~\cite{lal99,Spr08}. As the empirical information is
limited, it is desirable to compare the measured eigenvalue density
with a ``null hypothesis'' that results from a statistical ensemble.
The ensemble is defined~\cite{Mui82} by synthetic real or complex time
series $W_j, \ j=1,\ldots,p$ which yield the {\it empirical}
correlation matrix $C$ upon averaging over the probability density
function
\begin{equation}
P_{\beta}(W,C)\sim \exp \left(-\frac{\beta}{2} \tr W^{\dagger}C^{-1} W \right) \ ,
\label{eq2}
\end{equation}
that is, we have by construction
\begin{equation}
\int d[W] P_{\beta}(W,C)  \frac{1}{n} W W^\dagger = C \ ,
\label{eq3}
\end{equation}
where the measure $d[W]$ is the product of the differentials of all
independent elements in $W$. To ensure that $C$ is invertible, we
always assume $n\ge p$. When going to higher order statistics, the
Gaussian assumption~\eqref{eq2} is not necessarily justified, but it
often is a good approximation. This multivariate statistical approach
is closely related to Random Matrix Theory~\cite{GMW98}, and the
matrices $WW^{\dagger}$ are referred to as Wishart correlation
matrices.  One is interested in the ensemble averaged eigenvalue
density of these matrices. In terms of the resolvent, it reads
\begin{eqnarray}
 S_\beta(x)=-\frac{1}{p \pi} \im \int d[W] P_{\beta}(W,C) 
                 \tr\frac{\eins_p}{x^+\eins_p -WW^{\dagger}} ,
\label{2.3}
\end{eqnarray}
where $\eins_p$ is the $p \times p$ unit matrix.  The argument $x$
carries a small positive imaginary increment $\varepsilon>0$,
indicated by the notation $x^+=x+i\varepsilon$. The limit
$\varepsilon\to 0$ is implicit in our notation. Due to the invariance
of the trace and the measure, the ensemble averaged eigenvalue density
$S_\beta(x)$ only depends on the eigenvalues $\Lambda_j, \ j=1,\ldots ,p$ of $C$. Hence we may replace $C$ in~Eq.~\eqref{2.3} by the $p \times p$
diagonal matrix $\Lambda = \diag (\Lambda_1,\ldots, \Lambda_p)$. We
notice that the eigenvalues are positive definite, $\Lambda_j >0$.

A large body of literature is devoted to the eigenvalue
density~\eqref{2.3}.  Its asymptotic form for large $n$ and $p$ has
been studied in great detail, see Refs.~\cite{Sil95,VP2010}. However,
an exact closed--form result for finite $n$ and $p$ is only available
in the complex case~\cite{Alf04,SM04}.  Unfortunately, a deep,
structural mathematical reason made it up to now impossible to derive
such a closed--form result in the real case which is the more relevant
one for applications. We have three goals: We, first, introduce the
powerful supersymmetry method~\cite{Efe83,VWZ85,Efe97} to Wishart
correlation matrices for arbitrary $C$. This has, to the best of our
knowledge, not been done before. We, second, use the thereby achieved
unique structural clearness to derive a new and exact closed--form
result for the eigenvalue density in the real case for finite $n$ and
$p$. We, third, show that our results are easily numerically tractable
and compare them with Monte Carlo simulations.

Why does the real case pose such a substantial problem? --- This is
best seen by going to the polar decomposition $W=UwV$, where $U\in
{\rm O}(p) , \ V \in {\rm O}(n)$ for $\beta=1$ and $U\in {\rm U}(p) ,
\ V \in {\rm U}(n)$ for $\beta=2$ and where $w$ is the $p \times n$
matrix containing the singular values or radial coordinates $w_j, \
j=1,\ldots,p$. In particular, one has $WW^\dagger=Uw^2U^\dagger$, with $w^2=ww^{\dagger}$.
When inserting into~\eqref{2.3}, one sees that the non--trival group
integral
\begin{equation}
\Phi_{\beta}(\Lambda,w^2) = 
    \int \exp \left(-\frac{\beta}{2} \tr U^{\dagger}\Lambda^{-1} U w^2 \right) d\mu(U) 
\label{eq4}
\end{equation}
has to be done to obtain the joint probability density function of the
radial coordinates $w_j$. Here, $ d\mu(U)$ is the invariant Haar measure.
For $\beta=2$, this integral is the celebrated
Harish-Chandra--Itzykson--Zuber integral and known
explicitly~\cite{Har58,ItzZub80}. For $\beta=1$, however,
$\Phi_1(\Lambda,w^2)$, is not a Harish-Chandra spherical
function, it rather belongs to the Gelfand class~\cite{Gel50} and a
closed--form expression is lacking. The only explicit form known is a
cumbersome, multiple infinite series expansion in terms of zonal or Jack
polynomials~\cite{Mui82,Mac95}. This inconvenient feature then carries over
to the eigenvalue density~\eqref{2.3}, but we will arrive at
a finite series over twofold integrals. 

The supersymmetry method is based on writing
\begin{equation}
S_\beta (x) = -\frac{1}{ \pi p }\im\frac{\partial Z_{\beta}(J)}{\partial J} \Bigg|_{J=0} 
\label{2.5}
\end{equation}
as the derivative of the generating function
\begin{eqnarray}
Z_{\beta}(J) = \int d[W] P_{\beta}(W,\Lambda) \frac{\det( x^+
\eins_p + J \eins_p - WW^{\dagger} )}{\det( x^+ \eins_p 
- WW^{\dagger} )} 
\label{2.4}
\end{eqnarray}
with respect to the source variable $J$ at $J=0$. One has the
normaliziation $Z_{\beta}(0)=1$ at $J=0$. We consider the real and the
complex case and use the latter as test. We map $Z_{\beta}(J)$ onto
superspace using steps which are by now standard, see
Refs.~\cite{VWZ85,Efe97}.  A particularly handy approach for
applications such as the present one is given in Ref.~\cite{KGG08}, we
use the same conventions and find
\begin{eqnarray}
Z_{\beta}(J) &=& \int d[\rho] I_{\beta} (\rho)\nonumber\\
  &&\times
  \displaystyle{\prod_{j=1}^p {\detg}^{-\beta/2} \left(x^+\eins_{4/\beta} -J\gamma - \frac{\beta}{2}\Lambda_j \rho\right)}  . 
\label{3.1}
\end{eqnarray}
The merit of this transformation is the drastic reduction in the
number of degrees of freedom, because the variables to be intergrated
over form the $4/\beta \times 4/\beta$ Hermitean supermatrix
\begin{eqnarray}
\rho &=&   \begin{bmatrix}
     \rho_{11} & \rho_{12}^\dagger \\
     \rho_{12} & i\rho_{22}
\end{bmatrix}  \ .   
\label{eq5}
\end{eqnarray}
For $\beta=2$, $\rho_{11}$ and $\rho_{22}$ are scalar, real commuting
variables and $\rho_{12}$ is a complex anticommuting scalar variable.
For $\beta=1$, $\rho_{11}$ is a $2\times 2$ real symmetric matrix,
$\rho_{22}$ has to be multiplied with $\eins_2$ and we have
\begin{eqnarray}
\rho_{12}=\begin{bmatrix}
       \chi & \chi^*\\
    \xi & \xi^*
      \end{bmatrix} \ ,
\label{rho12}
\end{eqnarray}
where $\chi, \xi$ and $\chi^*, \xi^*$ denote anticommuting variables
and their complex conjugates, respectively.  We also introduced the
matrix $\gamma=\diag(0_{2/\beta} , - \eins_{2/\beta})$ and the
supersymmetric Ingham--Siegel integral
\begin{eqnarray}
I_{\beta}(\rho)= \int d[\sigma] \detg^{-n\beta/2}(\eins_{4/\beta}  +i\sigma) 
                                    \exp(i\trg \sigma\rho) \ ,
\label{3.2}
\end{eqnarray}
where $\sigma$ has the same form as $\rho$. The supertrace and
superdeterminant~\cite{Ber87} are denoted by $\trg$ and $\detg$.

Starting from the generating function~\eqref{3.1} we first consider
the complex case $\beta=2$. By introducing eigenvalue--angle
coordinates for the supermatrix $\rho$, we rederive in a
straightforward calculation the eigenvalue density $S_2(x)$ as found
in Ref.~\cite{Alf04}. In the real case $\beta=1$, the analogous
approach leads to inconvenient Efetov--Wegner terms~\cite{Efe97}, and
we thus proceed differently.  Since the generating function remains
invariant under rotations of the matrix $\rho_{11}$, we introduce its
eigenvalues $R_1 = \diag(r_1,r_2)$ and the diagonalizing angle as new
coordiantes. This yields the Jacobian $|\Delta_2(R_1)|=|r_1-r_2|$.
The next step is to evaluate the Ingham--Siegel integral $I_1(\rho)$.
The supermatrix $\sigma$ in Eq.~\eqref{3.2} has the same form as
$\rho$ in Eq.~\eqref{eq5}. Doing the integral over $\sigma$ followed
by an expansion in the anticommuting variables of $\rho$ according to
Eq.~\eqref{rho12} gives
\begin{eqnarray}
  && I_1(\rho) \sim {\det}^{(n-1)/2} R_1
  \exp\left(- \trg \rho\right) \Theta(R_1)\nonumber \\
  &&\quad\Biggl( \left(\frac{\partial}{i \partial \rho_{22}}\right)^{n-2}- 
                 \left(\frac{\chi \chi^*}{r_1} + \frac{\xi \xi^*}{r_2} \right)
                  \left(\frac{\partial}{i \partial \rho_{22}}\right)^{n-1} \nonumber \\
  &&\quad + \frac{1}{r_1r_2} \chi \chi^* \xi \xi^*
                     \left(\frac{\partial}{i \partial \rho_{22}}\right)^{n} \Biggr) \delta(\rho_{22}) \ .
\label{eq1.7}
\end{eqnarray}
In a simple, direct calculation, we also expand the product of the
superdeterminats in Eq.~\eqref{3.1} in the anticommuting variables of
$\rho$. We collect everything and do the integration over
the anticommuting variables. With the notation $Q_j=x^+ -2
\Lambda_ji\rho_{22}$, we obtain
\begin{eqnarray}
&&Z_1(J)  \sim \int d[R_1] \int d\rho_{22} |\Delta_2(R_1)| {\det}^{(n-1)/2}R_1 \Theta(R_1)   
           \nonumber \\
&&  \exp(-(r_1 +r_2 -2i\rho_{22}))  \prod_{j=1}^p  
          \frac{(J +Q_j) }{{\det}^{1/2} (x^+\eins_2 -2\Lambda_j r_1) }   \nonumber\\
&& \hspace*{-0.3cm}\Biggl( {\det}^{-1} R_1 
           \left( \frac{\partial}{i\partial  \rho_{22}}\right)^n 
                       +  \sum_{j=1}^p \frac{(2\Lambda_j)^2 }{(J +Q_j)} 
           \left( \frac{1}{( x^+- 2\Lambda_j r_1)r_2} \right.      \nonumber \\
&&      \left. +\frac{1}{( x^+ - 2\Lambda_j r_2)r_1}\right) 
                \left( \frac{\partial}{i\partial  \rho_{22}}\right)^{n-1}   \nonumber\\
&&        +\sum_{j\neq k}^p \left(\frac{(2\Lambda_j)^2 }
                                    {(J +Q_j)( x^+ - 2\Lambda_j r_1)}\nonumber \right. \\
&&   \left.  \hspace*{0cm} \times\frac{(2\Lambda_k)^2}
              { (J+Q_k) ( x^+ - 2\Lambda_k r_2)} 
                     \left( \frac{\partial}{i\partial  \rho_{22}}\right)^{n-2}\right) \Biggr) \delta(\rho_{22}) \ .
\label{eq1.9}
\end{eqnarray}
According to Eq.~\eqref{2.5} we have to take the derivative with respect to
$J$. This leads to the three relations
\begin{eqnarray}
&&\frac{\partial}{\partial J} \prod_{l=1}^p(J+Q_l)\Bigg|_{J=0}  =E_{p-1}(Q)\\
&&\frac{\partial}{\partial J}\frac{1}{J +Q_j} \prod_{l=1}^p(J+Q_l)\Bigg|_{J=0}= E_{p-2;j}(Q)\nonumber\\
&&\frac{\partial}{\partial J}\frac{1}{J +Q_j}\frac{1}{J +Q_k} \prod_{l=1}^p(J+Q_l)\Bigg|_{J=0} = E_{p-3;j,k}(Q) \ , \nonumber
\end{eqnarray}
where $E_{m; i,k}(Q)$ denotes the elementary symmetric polynomial of
order $m$ in the variables $Q_j,\ j=1,\ldots,p \neq i,k$ with $Q_i$
and $Q_k$ omitted, 
\begin{eqnarray}
 E_{m; i,k}(Q) = \sum_{\underset{{\neq i \neq k}}{1\leq i_1 <
 \cdots <i_{m} \leq p}} Q_{i_{1}}\ldots Q_{i_{m}} \ .
\end{eqnarray}
We finally arrive at 
\begin{eqnarray}
&&\hspace*{-0.3cm}S_{1} (x)=c\,\im\int d[R_1] \int d\rho_{22}   
   \frac{ |\Delta_2(R_1)| {\det}^{(n-1)/2} R_1 }{ \prod_{j=1}^p\det^{1/2} ( x^+\eins_2 - 2\Lambda_j R_1)}\nonumber\\
&&\Theta(R_1)e^{-(r_1 +r_2-2i\rho_{22})} \Biggl({\det}^{-1} R_1 E_{p-1}(Q) \left( \frac{\partial}{i\partial  \rho_{22}}\right)^n  \nonumber \\
&&+\sum_{j=1}^p (2\Lambda_j)^2\left( \frac{1}{( x^+  - 2\Lambda_j r_2)r_1 } \right. \label{eq1.10}\\
&&\left.\hspace*{1.7cm} +\frac{1}{( x^+  - 2\Lambda_j r_1)r_2 }\right)E_{p;j}(Q) \left( \frac{\partial}{i\partial  \rho_{22}}\right)^{n-1} \nonumber   \\    
&&\hspace*{-0.3cm}+\sum_{j\neq k}^p \frac{(2\Lambda_j)^2(2 \Lambda_k)^2E_{p-3,j,k}(Q) }{(x^+ -2\Lambda_j r_1)(x^+-2\Lambda_kr_2)}  \left( \frac{\partial}{i\partial  \rho_{22}}\right)^{n-2} \Biggr) \delta(\rho_{22}) \ , \nonumber
\end{eqnarray}
where the constant reads $c=(-1)^{n+1}/(4 \pi p (n-2)!)$.
Due to the $\delta$--distribution, the integral over $\rho_{22}$ are
elementary. Hence we end up with an expression for the eigenvalue
density $S_1(x)$ which essentially is a twofold integral.

The integrals in Eq.~\eqref{eq1.10} can be numerically evaluated by
using a regularisation technique of the type
\begin{eqnarray}
&&   \im
\int\limits_{0}^{\infty}dr_1\int\limits_{0}^{\infty}dr_2\frac{f(r_1,r_2)}
{\prod\limits_{l=1}^p\sqrt{r_1 -\Lambda_l^{-1}}\sqrt{r_2
-\Lambda_l^{-1}}}=\nonumber\\
&=& \sum_{\substack{ 0\leq i,j \leq p\\ (i+j)\in 2\mathbb{N} +1}}\int\limits_{\Lambda_{i+1}^{-1}}^
{\Lambda_{i}^{-1}}dr_1\int\limits_{\Lambda_{j+1}^{-1}}^{\Lambda_{j}^{-1}}dr_2f(r_1,r_2)\nonumber\\
&&\hspace*{2cm}\frac{1}{\prod\limits_{l=1}^p\sqrt{r_1 -\Lambda_l^{-1}}\sqrt{r_2
-\Lambda_l^{-1}}} \ .
\label{5.8} 
\end{eqnarray}
Here we assume an ordering of the eigenvalues such that $
\Lambda_0>\Lambda_1 > \ldots > \Lambda_p>\Lambda_{p+1}$ with
$\Lambda_0^{-1}=0$ and $\Lambda_{p+1}^{-1}=\infty$. The real function
$f(r_1,r_2)$ is independent of $\varepsilon$ and has no singularities.
The singularties at the boundaries of the domain are integrable. Using
the commercial software \textsc{Mathematica}\textsuperscript{\textregistered}  ~\cite{math}, we evaluate 
our formula~\eqref{5.8} numerically. For independent comparison, we
also carry out Monte Carlo simulations with ensembles of $10^5$ matrices. 
In Figs.~\ref{fig1} and \ref{fig2}, we show the
\begin{widetext}
\begin{center}
\begin{figure}[t!]
\includegraphics[scale=0.5]{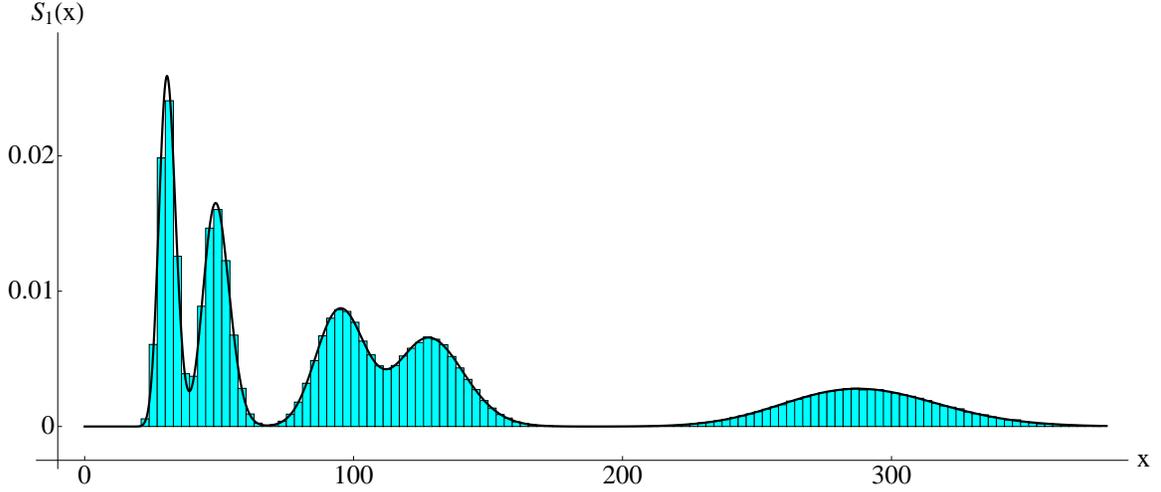}
\caption{Eigenvalue density for $p=5$ and $n=200$: analytical
  formula (solid lines) and Monte Carlo simulations (histogram with bin width 3).}
\label{fig1}
\end{figure}
\end{center}
\end{widetext}
\begin{widetext}
\begin{center}
\begin{figure}[t!]
\includegraphics[scale=0.63]{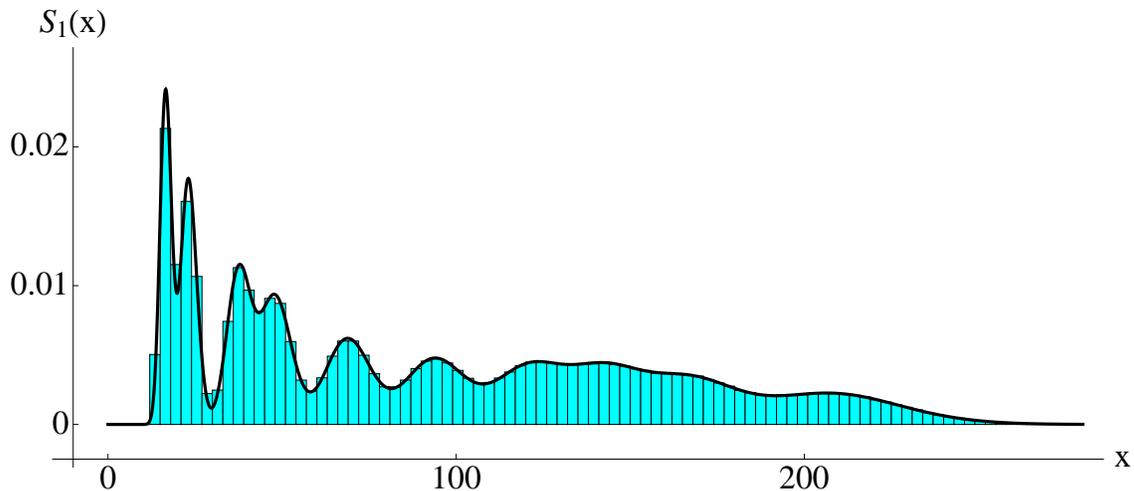}
\caption{Eigenvalue density for $p=10$ and $n=200$: analytical
  formula (solid lines) and Monte Carlo simulations (histogram with bin width 3).}
\label{fig2}
\end{figure}
\end{center}
\end{widetext}
results for $p=5$ and $p=10$ and $n=200$ with the chosen empirical eigenvalues
$\Lambda_j$, $j=1,\ldots,5$ of $\{1.44, 0.64, 0.49, 0.25, 0.16    \}$ and
$\Lambda_j$, $j=1,\ldots,10$ of $\{1, 0.81, 0.7225, 0.64, 0.45, 0.36, 0.25, 0.2025, 0.1225, 0.03\}$, respectively. The agreement is perfect.

In conclusion, we introduced the supersymmetry method for the first
time to Wishart correlation matrices. We thereby derived exact
expressions for the eigenvalue density in terms of low--dimensional
integrals.  This is a drastic reduction, as the original order of
integrals is $np$.  Our approach solves a serious mathematical
obstacle in the real case.  A presentation for a mathematics audience
will be given elsewhere~\cite{RKG10}. Here, we derived and discussed
the formulae needed for applications. In the real case ($\beta=1$), we
obtained the previously unknown exact solution in terms of a finite sum
of twofold integrals. We evaluated our formula numerically and
confirmed it by comparing to Monte Carlo simulations.

We thank R.~Sprik for fruitful discussions as well as A.~Hucht,
H.~Kohler, and R.~Sch\"afer for helpful comments. One of us (TG)
greatly benefitted from the Program on High Dimensional Inference and
Random Matrices in 2006 at SAMSI, Research Triangle Park, North
Carolina (USA).  We acknowledge support from Deutsche
Forschungsgemeinschaft (Sonderforschungsbereich Transregio 12).

%%\bibliographystyle{unsrt}
%%\bibliography{Referenzen}

\end{document}